\documentclass[twocolumn,showpacs,preprintnumbers,amsmath,amssymb]{revtex4-1}
\usepackage{bbm}
\usepackage{amsfonts}

\usepackage{epsfig}
\usepackage{graphicx}
\usepackage{amssymb}   
\usepackage{dcolumn}
\usepackage{bm}
\usepackage[colorlinks,citecolor=blue, linkcolor=blue,hyperindex,CJKbookmarks]{hyperref}
\begin{document}
\vspace{2.5cm}
\title{Enhancing teleportation of quantum Fisher information by partial measurements}
\author{Xing Xiao$^{1}$}
\author{Yao Yao$^{2}$}
\author{Wo-Jun Zhong$^{1}$}
\author{Yan-Ling Li$^{3}$}
\author{Ying-Mao Xie$^{1}$}
\altaffiliation{xieyingmao@126.com}
\affiliation{$^{1}$College of Physics and Electronic Information,
Gannan Normal University, Ganzhou Jiangxi 341000, China\\
$^{2}$Microsystems and Terahertz Research Center, China Academy of Engineering Physics, Chengdu, Sichuan 610200, China\\
$^{3}$School of Information Engineering, Jiangxi University of Science and Technology, Ganzhou, Jiangxi 341000, China\\}

\begin{abstract}
The purport of quantum teleportation is to completely transfer information 
from one party to another distant partner. However, from the perspective of parameter estimation, it is the information carried by a particular parameter, not the information of total quantum state that needs to be teleported. Due to the inevitable noise in environment, we propose two schemes to enhance quantum Fisher information (QFI) teleportation under amplitude damping noise with the technique of partial measurements. We find that post partial measurement can greatly enhance the teleported QFI, while the combination of prior partial measurement and post partial measurement reversal could completely eliminate the effect of decoherence.
We show that, somewhat consequentially, enhancing QFI teleportation is more economic than that of improving fidelity teleportation. Our work extends the ability of partial measurements as a quantum technique to battle decoherence in quantum information processing.

\end{abstract}
\maketitle

\section{Introduction}
\label{intro}

Quantum teleportation, one of the most fascinating protocols predicted by quantum mechanics \cite{bennett93},
is a critical ingredient for quantum communication and quantum computation networks \cite{gottesman99,ursin04}. 
It is the faithful transfer of quantum states between two distant parties which have
established prior entanglement and can communicate classically. In the past two decades, quantum
teleportation has attracted considerable attentions and been studied both theoretically and experimentally \cite{braunstein98,bouwmeester97,riebe04,barret04,olms09,pfaff14,wang15,takesue15}.
However, in many scenarios, it is not the whole quantum state, but rather the information of a particular parameter physically encoded in it, that is needed to be transmitted. Therefore, there is no need to teleport the full information of the quantum states themselves, but only the relevant parameter information is our practical concern. In contrast to the quantum state teleportation where the credibility of teleportation is measured by fidelity, the transmission of information that carried by a physical parameter is usually quantified by quantum Fisher information (QFI) \cite{helstrom76,holevo82,kay93,genoni11,lu13}. QFI plays a significant role in the fields of quantum geometry of state spaces \cite{helstrom76,holevo82,wootters81}, quantum information theory \cite{nielsen00} and quantum metrology \cite{giov06,giov11}. Particularly, the inverse of QFI characterizes the ultimate achievable precision in parameter estimation \cite{braunstein94}.

Unfortunately, any realistic quantum system inevitably couples to
other uncontrollable environments which influence it in a non-negligible way \cite{breuer1}.
Then the issue of robustness of QFI against various sources of decoherence has soon been raised. Numerous researches have
indeed demonstrated that QFI is fragile and easily broken by environmental noise \cite{kolo10,ma11,berrada12,zhang13,liyanling15}. This would be the most limiting factor for the applications of QFI in quantum teleportation, quantum metrology and other quantum tasks.
In this context, it is an extremely important issue to protect the QFI from decoherence during the procedure of teleportation, especially for the atomic, trapped ions and other solid-state systems \cite{riebe04,barret04,olms09,pfaff14}. 

Partial measurements, which are generalizations of von
Neumann measurements, are associated with positive-operator valued measures (POVM). For partial measurements \cite{korotkov99,korotkov06,korotkov10,para11a,para11b}, the information extracted from the quantum state is deliberately limited, thereby keeping the measured state alive (i.e.,  without completely collapsing towards an eigenstate). Thus, it would be possible to retrieve the initial information with some operations even when the quantum state has suffered decoherence. Recently, many proposals that exploit partial measurements to protect the fidelity of a single qubit, the quantum entanglement of two qubits and two qutrits from AD decoherence have been demonstrated in both theoretically \cite{sun09,man12,xiao13} and experimentally \cite{katz08,kim09,kim12}. This motivates us to study the QFI teleportation under decoherence by utilizing the partial measurements.

In this paper, we  propose two schemes to show that partial measurements can greatly enhance the QFI teleportation under decoherence. In particular, we find that the combination of prior partial measurement and post partial measurement reversal is able to completely circumvent the influence of AD noise. Our schemes for enhancing QFI are based on the fact that partial measurements are nondestructive and can be reversed with a certain probability. Moreover, we analytically obtain the optimal parameters of the enhancement of QFI teleportation. We also demonstrate that the success probability of enhancing the QFI is higher than that of enhancing the fidelity of quantum state, which indicates that enhancing the teleported QFI is a more reasonable and economic way in the scenario of parameter estimation.

This paper is organized as follows: we introduce QFI and partial measurements in Sec. \ref{sec:2}. In Sec. \ref{sec:3}, we then show how the QFI teleportation could be enhanced by
partial measurements. We consider two different protocols, as illustrated in Fig. \ref{Fig1}, and compare the results in Refs. \cite{prama13,qiu14}  where the fidelity of teleportation is improved by partial measurements. We show that the cost of enhancing QFI is smaller than that of improving fidelity, which is represented as the higher success probability. Finally, the conclusions are summarized in Sec. \ref{sec:4}.

\section{Preliminaries}
\label{sec:2} 
\subsection{Quantum Fisher Information}
QFI of a parameter characterizes the sensitivity of the state with respect to changes
of the parameter. We consider that a parameter $\phi$ is encoded in quantum states $\rho_{\phi}$ (in general, mixed states). The QFI of $\phi$ is defined as 
\cite{helstrom76,holevo82,braunstein94}
\begin{equation}
\label{e1}
\mathcal {F}_{\phi}=\textrm{Tr}(\rho_{\phi}\mathcal {L}_{\phi}^{2})=\textrm{Tr}[(\partial_{\phi} \rho_{\phi})L_{\phi}],
\end{equation}
where $\mathcal {L}_{\phi}$ is the so-called symmetric logarithmic
derivative, which is defined by
$\partial_{\phi}\rho_{\phi}=(\mathcal
{L}_{\phi}\rho_{\phi}+\rho_{\phi}\mathcal {L}_{\phi})/2$
with $\partial_{\phi}=\partial/\partial\phi$. Typically, there are three methods to calculate the QFI \cite{yao14}. The most frequently used one is diagonalizing
the matrix as
$\rho_{\phi}=\Sigma_{n}\lambda_{n}|\psi_{n}\rangle\langle\psi_{n}|$.
Then one can rewritten the QFI as \cite{knysh11,liu13}
\begin{equation}
\label{e2}
\mathcal {F}_{\phi}=\sum_{n}\frac{(\partial_{\phi}\lambda_{n})^2}{\lambda_{n}}+\sum_{n}\lambda_{n}\mathcal{F}_{\phi,n}
-\sum_{n\neq m}\frac{8\lambda_{n}\lambda_{m}}{\lambda_n+\lambda_m}|\langle\psi_{n}|\partial_{\phi}\psi_{m}\rangle|^2,
\end{equation}
where $\mathcal {F}_{\phi,n}$ is the QFI for pure state $|\psi_{n}\rangle$ with the form
$\mathcal {F}_{\phi,n}=4[\langle\partial_{\phi}\psi_{n}|\partial_{\phi}\psi_{n}\rangle-|\langle\psi_{n}|\partial_{\phi}\psi_{n}\rangle|^2]$. According to Eq. (\ref{e2}), the last term stemming from the mixture of pure states suggests that the QFI of a mixed state is smaller then pure-state case. 

For the single qubit state, a simple and
explicit expression of QFI could be obtained. In the Bloch sphere
representation, any qubit state can be written as
\begin{equation}
\label{e3}
\rho = \frac{1}{2}(1+\vec{r}\cdot\hat{\sigma}),
\end{equation}
where $\vec{r}=(r_{x}, r_{y}, r_{z})^{T}$ is the real Bloch vector
and $\hat{\sigma}=(\hat{\sigma}_{x}, \hat{\sigma}_{y},
\hat{\sigma}_{z})$ denotes the Pauli matrices. Therefore, for the
single qubit state, $\mathcal{F}_{\phi}$ can be represented as follows \cite{zhong13}
\begin{eqnarray}
\mathcal{F}_{\phi}=\left\{\begin{array}{cc}
|\partial_{\phi} \vec{r}|^{2}+\frac{\vec{r}\cdot\partial_{\phi}\vec{r}}{1-|\vec{r}|^{2}}, & \mbox{ if } \, |\vec{r}|<1,\\
|\partial_{\phi}\vec{r}|^{2}, & \mbox{ if } \, |\vec{r}|=1.
\end{array}\right.
\label{e4}
\end{eqnarray}

\begin{figure}
  \includegraphics[width=0.5\textwidth]{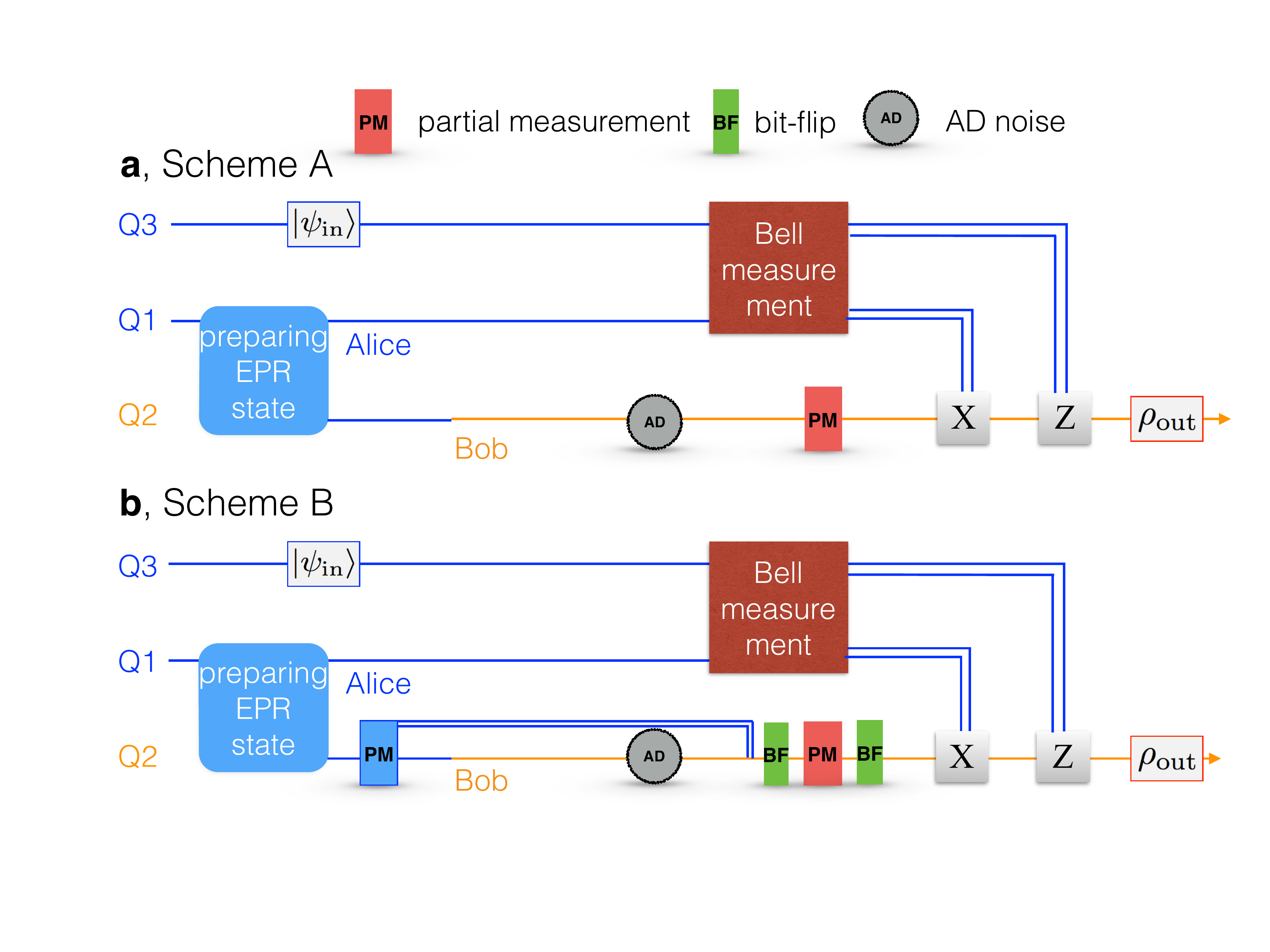}
\caption{(color online) Schematic illustrations of enhancing QFI teleportation under decoherence using partial measurements. (a) The circuit of scheme $A$ is similar to the standard circuit of quantum teleportation while a post partial measurement is added before the unitary operations. (b) In the circuit of scheme $B$, Alice performs a prior partial measurement (with strength $p$) on the 2nd qubit before she sends it to Bob. During their classical communications, Alice has to send the results of Bell measurement and the prior partial measurement strength $p$ to Bob. When Bob receives the results, he firstly performs a post partial measurement reversal (i.e., bit-flip, partial measurement and bit-flip) and then sequentially performs the local unitary operations.}
\label{Fig1}       
\end{figure}

\subsection{Partial Measurements}
In quantum physics, the standard von Neumann projective measurements are referred to as ``sharp measurements'', which project the initial state to one of the eigenstates of the measurement operator. As generalizations of the standard von Neumann projective measurements, partial measurements don't completely collapse the initial state  
(i.e., non-projective measurements), hence they have the interesting property that they can be reversed in a probabilistic way. For a single qubit with computational basis $|0\rangle$ and $|1\rangle$, the so-called partial measurement is 
\begin{eqnarray}
\label{e5}
\mathcal{M}_{0}&=&|0\rangle\langle0|+\sqrt{1-p}|1\rangle\langle1|,\\
\label{e6}
\mathcal{M}_{1}&=&\sqrt{p}|1\rangle\langle1|,
\end{eqnarray}
where parameter $p$ $(0\leqslant p\leqslant1)$ is usually named as the strength of partial measurement. Note that,  $\mathcal{M}_{0}$ and $\mathcal{M}_{1}$ are not necessarily projectors and also nonorthogonal to each other, but $\mathcal{M}_{0}^{\dagger}\mathcal{M}_{0}+\mathcal{M}_{1}^{\dagger}\mathcal{M}_{1}=I$.  Though the measurement operator $\mathcal{M}_{1}$ is the same as von Neumann projective measurement and associated with irrevocable collapse,  $\mathcal{M}_{0}$ is a partial measurement which we focus in this paper.

\subsection{Partial Measurements Reversal}
The partial measurement $\mathcal{M}_{0}$ has some interesting properties: (i) the strength $p$ is controllable and (ii) it could be reversed for the case $p\neq1$.
The reverse procedure can be noticed immediately to be
\begin{eqnarray}
\label{e7}
\mathcal{M}_{0}^{-1}&=&|0\rangle\langle0|+\frac{1}{\sqrt{1-p}}|1\rangle\langle1|, \\
&=&\frac{1}{\sqrt{1-p}}X\mathcal{M}_{0}X,\nonumber
\end{eqnarray}
where $X=|0\rangle\langle1|+|1\rangle\langle0|$ is the qubit bit-flip operation. The second line of Eq. (\ref{e7}) indicates that the reverse process $\mathcal{M}_{0}^{-1}$ can be achieved physically by three sequential operations: bit-flip, partial measurement and bit-flip.

\section{Enhancing QFI teleportation by partial measurements}
\label{sec:3}
In realistic quantum teleportation, the maximally entangled state may lose its coherence and become a mixed state due to the interaction with its environment. 
Noted that the quantum channel
which is less entangled will reduce the teleported QFI. Here, we propose two
schemes to enhance the QFI teleportation under amplitude damping (AD) noise, as shown in Fig. \ref{Fig1}. For simplicity, we consider the scenario that Alice has a perfect quantum memory while Bob doesn't. Alice prepares the EPR state and then sends one particle to Bob. Therefore, the entangled part possessed by Bob will suffer AD noise and hence reduces the entanglement. The situation that both Alice and Bob influenced by noises could be treated similarly.

\subsection{Scheme $A$}

We first examine the efficiency of scheme $A$, in which only one post partial measurement
is performed by Bob before local operations (e.g., $X$ or $Z$). We assume the shared EPR state is prepared in $|\Psi^{+}\rangle=(|00\rangle+|11\rangle)/\sqrt{2}$ by Alice and the input quantum state is an arbitrary
superposition state
\begin{equation}
\label{e8}
|\psi_{\rm in}\rangle=\cos\frac{\theta}{2}|0\rangle+\sin\frac{\theta}{2}e^{i\phi}|1\rangle
\end{equation}
which carries a phase parameter $\phi$. Note that the QFI of $\phi$ is our concern during this teleportation procedure. The input state of the quantum teleportation circuit in
Fig. \ref{Fig1}a is the product state of $|\psi_{\rm in}\rangle$ and $|\Psi^{+}\rangle$.

While the EPR state is
prepared and stored perfectly by Alice, the qubit kept by Bob 
may be affected by AD noise. The
dynamics of an entangled state subject to AD noise
is described by the quantum operation $\Lambda$ acting on the pure state \cite{nielsen00} 
\begin{eqnarray}
\rho^{AD}&=&\Lambda(\rho_{0})=\sum_{i=1,2}E_{i}|\Psi^{+}\rangle\langle\Psi^{+}|E_{i}^{\dagger},\\
&=&\frac{1}{2}(|00\rangle\langle00|+\gamma|10\rangle\langle10|+\overline{\gamma}|11\rangle\langle11|\nonumber\\
&&+\sqrt{\overline{\gamma}}|00\rangle\langle11|+\sqrt{\overline{\gamma}}|11\rangle\langle00|),\nonumber
\label{e9}
\end{eqnarray}
where $\rho_{0}=|\Psi^{+}\rangle\langle\Psi^{+}|$ and the subscript $AD$ denotes the pure AD noise case (i.e., without partial measurements).  $E_{i}$ are the Kraus operators of AD noise
\begin{eqnarray}
E_{1}=I_{1}\otimes\left(\begin{array}{cc}1 & 0 \\0 & \sqrt{1-\gamma}\end{array}\right), 
E_{2}=I_{1}\otimes\left(\begin{array}{cc}0 & \sqrt{\gamma} \\0 & 0\end{array}\right),
\label{e10}
\end{eqnarray}
with $I_{1}$ is the two dimensional identity operator of the 1st qubit (Q1) since we have assumed Alice is not affected by AD noise. Note that following the standard circuit of quantum teleportation process, we get the teleported QFI under AD noise 
\begin{equation}
\mathcal{F}_{\phi}^{AD}=\sin^2\theta(1-\gamma).
\label{e11}
\end{equation}

Then we consider the effect of a post partial measurement (with the strength $p_{\rm r}^{A}$) performed by Bob before he makes the corresponding unitary operations. Here, the superscript $A$ ($B$) indicates it belongs to the scheme $A$ ($B$).
\begin{eqnarray}
\label{ea1}
\rho^{A}&=&M_{0}\left[\Lambda(\rho_{0})\right]M_{0}^{\dagger}\\
&=&\frac{2}{1+\overline{p}_{\rm r}^{A}}(|00\rangle\langle00|+\overline{p}_{\rm r}^{A}\gamma|10\rangle\langle10|+\overline{p}_{\rm r}^{A}\overline{\gamma}|11\rangle\langle11|\nonumber\\
&&+\sqrt{\overline{p}_{\rm r}^{A}\overline{\gamma}}|00\rangle\langle11|+\sqrt{\overline{p}_{\rm r}^{A}\overline{\gamma}}|11\rangle\langle00|),\nonumber
\end{eqnarray}
where $M_{0}=I_{1}\otimes\mathcal{M}_{0}$. After the implementation of the quantum circuit of Fig. \ref{Fig1}a, Bob 
gets the teleported state $\rho_{\rm out}^{A}$ whose three Bloch vector components are
\begin{eqnarray}
r_{x}^{A}&=&\frac{2\sin\theta\cos\phi}{N^{A}}\sqrt{\overline{p}_{\rm r}^{A}\overline{\gamma}},\\
\label{e12}
r_{y}^{A}&=&\frac{2\sin\theta\sin\phi}{N^{A}}\sqrt{\overline{p}_{\rm r}^{A}\overline{\gamma}},\\
\label{e13}
r_{z}^{A}&=&\frac{\cos\theta}{N^{A}}(1+\overline{p}_{\rm r}^{A})\overline{\gamma},
\label{e14}
\end{eqnarray}
where $p_{\rm r}^{A}$ is the strength of partial measurement and $\overline{p}_{\rm r}^{A}=1-p_{\rm r}^{A}$, $\overline{\gamma}=1-\gamma$. $N^{A}=2-p_{\rm r}^{A}-\gamma p_{\rm r}^{A}$ is the normalized
factor. According to Eq. (\ref{e4}), the QFI of teleported state could be obtained
\begin{equation}
\mathcal{F}_{\phi}^{A}=\frac{4\sin^2\theta\overline{p}_{\rm r}^{A}\overline{\gamma}}{(N^{A})^2}.
\label{e15}
\end{equation}
Since the strength of partial measurement is tunable, we can maximize the QFI by choosing the optimal partial measurement strength
\begin{equation}
p_{\rm r}^{A,\rm opt}=2\gamma/(1+\gamma).
\label{e16}
\end{equation}
Then the maximal QFI is 
\begin{equation}
\mathcal{F}_{\phi}^{A,\rm opt}=\frac{\sin^2\theta}{1+\gamma}.
\label{e17}
\end{equation}

\begin{figure}
  \includegraphics[width=0.5\textwidth]{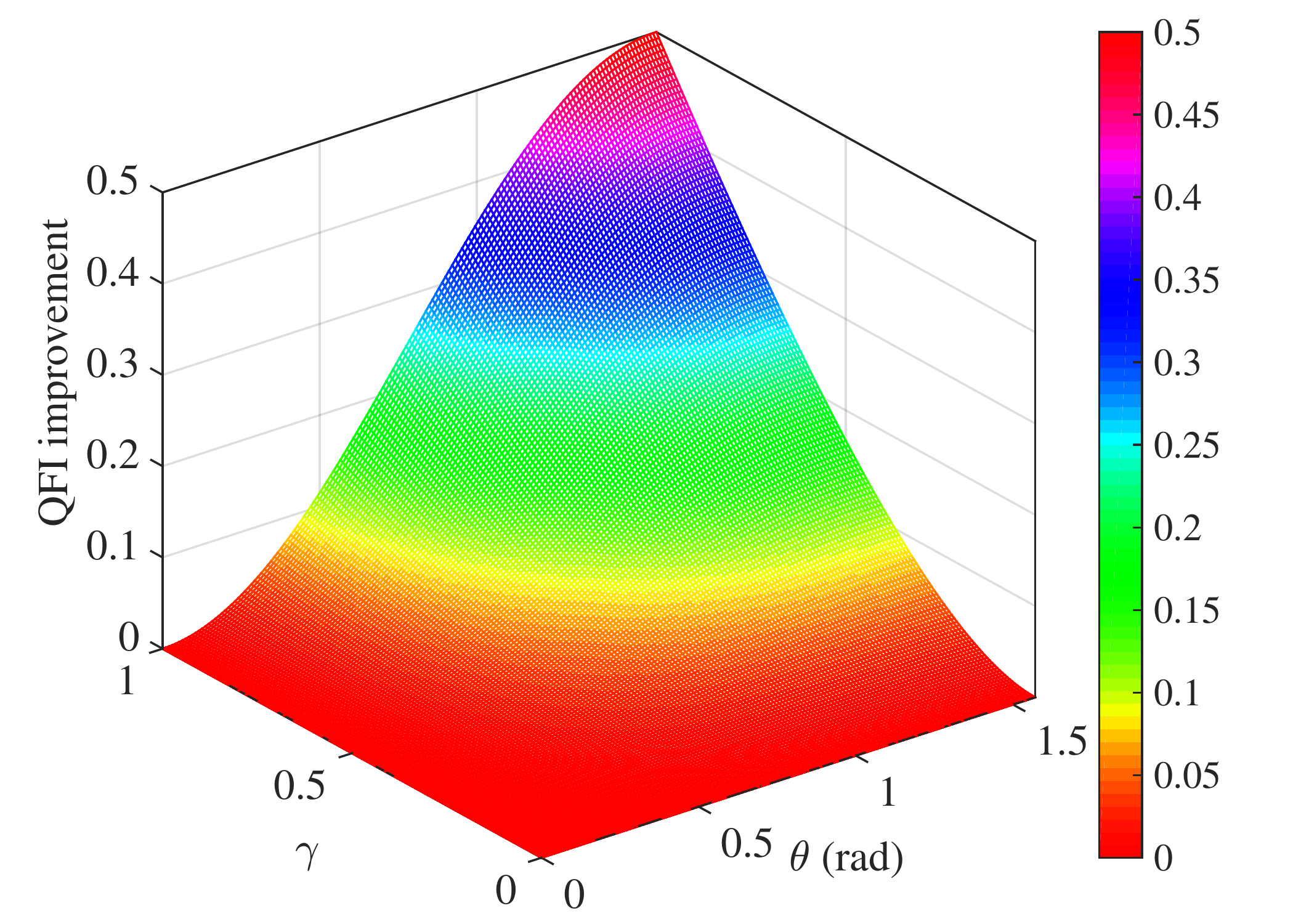}
\caption{(color online) The improved QFI $\mathcal{F}_{\rm imp}^{A}$ by post partial measurement as a function of $\theta$ and dimensionless parameter $\gamma$.}
\label{Fig2}       
\end{figure}

To quantify the efficiency of partial measurement on enhancing QFI, we introduce the teleported QFI improvement,
\begin{equation}
\mathcal{F}_{\rm imp}^{A}\equiv\mathcal{F}_{\phi}^{A,\rm opt}-\mathcal{F}_{\phi}^{\rm AD},
\label{e18}
\end{equation}
which is plotted in Fig. \ref{Fig2} as a function of decoherence strength $\gamma$ and initial parameter $\theta$.
It is remarkable to find that the teleported QFI improvement $\mathcal{F}_{\rm imp}^{A}$ is always non-negative. Particularly, when $\gamma=1$, we find $\mathcal{F}_{\phi}^{\rm AD}=0$ while $\mathcal{F}_{\phi}^{A,\rm opt}=\sin^2\theta/2$, which means that half of the QFI is still teleported under complete decoherence with the help of post partial measurement.
Our result indicates that the application of quantum partial measurement indeed enhances the QFI teleportation, which is very important for quantum metrology and other quantum information tasks.

The underlying physical mechanism of the enhancement of QFI needs to be clarified further. One might deduce that the enhancement of teleported QFI should be attributed to the improvement of entanglement between Alice and Bob. We argue that this is not the case. 
According to Eqs. (\ref{e9}) and (\ref{ea1}) and following the definition of concurrence \cite{wootters98}, we can obtain the concurrence of $\rho^{AD}$ and $\rho^{A}$, respectively, which are $C^{AD}=\sqrt{\overline{\gamma}}$ and $C^{A}=2\sqrt{\overline{p}_{\rm r}^{A}\overline{\gamma}}/(1+\overline{p}_{\rm r}^{A})$. Employing the optimal partial measurement strength of QFI, i.e., Eq. (\ref{e16}), we have $C^{A,\rm opt}=\overline{\gamma}\sqrt{1+\gamma}$. However, it is contrary to one's expectation that $C^{A,\rm opt}/C^{AD}=\sqrt{1-\gamma^{2}}\leqslant1$, which means that, on the optimal condition of teleported QFI, the entanglement between Alice of Bob is not improved with the help of post partial measurement.  Therefore, the enhancement of QFI in scheme $A$ could not be attributed to the improvement of entanglement, but rather to the probabilistic nature of partial measurement. 

In contrast to the results in Refs. \cite{prama13,qiu14}, our ultimate aim is not to enhance
the whole state fidelity but rather to improve the information of a particular parameter encoded in the teleported state (i.e., the QFI). Therefore, as one might expect, improving the QFI teleportation is easier than enhancing the fidelity. 
This intuition could be confirmed by comparing the success probability. In the Refs. \cite{prama13,qiu14}, the success probability of improving the fidelity is
$P_{\rm fid}^{A}=1-\frac{\gamma(3+\gamma)}{2(1+\gamma)}$, while in our scenario the success probability of enhancing QFI is
\begin{equation}
P_{\rm QFI}^{A}=1-\gamma.
\label{e19}
\end{equation}
Similarly, we define the success probability improvement $P_{\rm imp}^{A}$ and it is easy to check that 
\begin{equation}
P_{\rm imp}^{A}\equiv P_{\rm QFI}^{A}-P_{\rm fid}^{A}=\frac{\gamma(1-\gamma)}{2(1+\gamma)}\geqslant0,
\label{e20}
\end{equation} 
as $\gamma\in[0,1]$.  This means if we focus on teleporting the information of a particular parameter, we have no need to improve the teleportation fidelity of the whole state. Enhancing the QFI teleportation would be more reasonable and economic in this case.

\subsection{Scheme $B$}
Although a post partial measurement can enhancing the QFI teleportation, part of QFI is still lost in the decoherence process. This is not our ultimate purpose. In this section, we show that an improved scheme can completely circumvent the decoherence and retrieve
all the initial QFI. The key improvement is that Alice performs a prior partial measurement (with the measurement strength $p$) on the second qubit (Q2) before she sends it to Bob. During their classical communications, Alice has to send the results of Bell measurement 
and the information of $p$ to Bob. When Bob receives the results, he firstly carries out a post partial measurement reversal (i.e., bit-flip, partial measurement and bit-flip) and then sequentially performs the local unitary operations. Remarkably, we show below that the prior partial measurement plays a significant role in enhancing the QFI teleportation.

\begin{figure*}
  \includegraphics[width=0.9\textwidth]{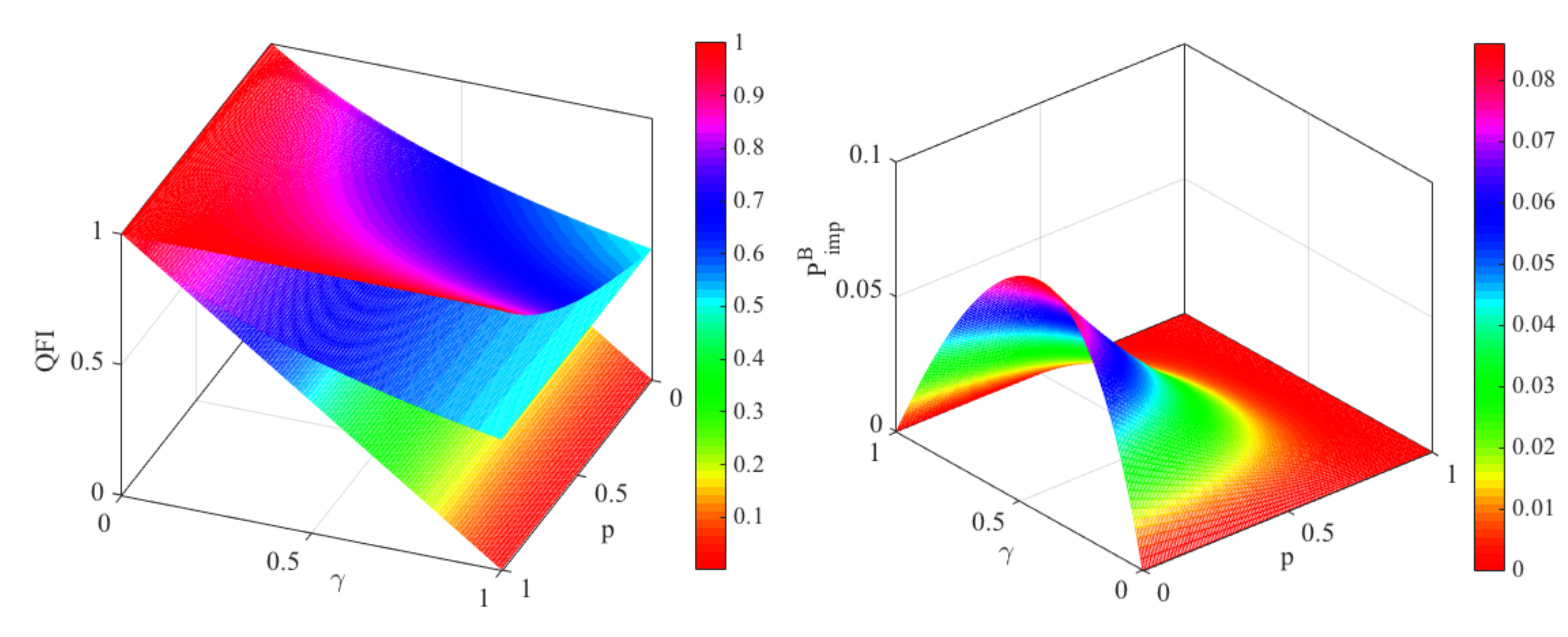}
\caption{(color online) (a). Teleported QFI $\mathcal{F}_{\phi}^{B,\rm opt}$,  $\mathcal{F}_{\phi}^{A,\rm opt}$ and $\mathcal{F}_{\phi}^{AD}$ as a function of dimensionless parameters $p$ and $\gamma$. The three surfaces from top to bottom correspond to scheme $B$, scheme $A$ and pure AD noise cases, respectively. Note that $\mathcal{F}_{\phi}^{A,\rm opt}$ and $\mathcal{F}_{\phi}^{AD}$ are independent of $p$ so they are planes. The parameter $\theta$  has been set to $\pi/2$. (b). The success probability improvement $P_{\rm imp}^{B}$ as a function of dimensionless parameters $p$ and $\gamma$.}
\label{Fig3}       
\end{figure*}

According to the circuit depicted in Fig. \ref{Fig1}b, the second qubit (Q2) experiences three processes: prior partial measurement, AD noise and post partial measurement reversal. The final entangled state could be represented as
$\rho^{B}=M_{0}^{-1}\left[\Lambda\left(M_{0}\rho_{0}M_{0}^{\dagger}\right)\right](M_{0}^{-1})^{\dagger}$,
where $M_{0}^{-1}=I_{1}\otimes\mathcal{M}_{0}^{-1}$. The following QFI teleportation is based on above entangled state and the finally teleported state is $\rho_{\rm out}^{B}$. The corresponding three Bloch vector components are
\begin{eqnarray}
r_{x}^{B}&=&\frac{2\sin\theta\cos\phi}{N^{B}}\sqrt{\overline{p}\ \overline{p}_{\rm r}^{B}\overline{\gamma}},\\
\label{e21}
r_{y}^{B}&=&\frac{2\sin\theta\sin\phi}{N^{B}}\sqrt{\overline{p}\ \overline{p}_{\rm r}^{B}\overline{\gamma}},\\
\label{e22}
r_{z}^{B}&=&\frac{\cos\theta}{N^{B}}(\overline{p}_{\rm r}^{B}\overline{\gamma}+\overline{p}\ \overline{\gamma}+p\gamma\overline{p}_{\rm r}^{B}),
\label{e23}
\end{eqnarray}
with $\overline{p}=1-p$ and $N^{B}=\overline{p}_{\rm r}^{B}+\overline{p}\ \overline{\gamma}+\overline{p}\gamma\overline{p}_{\rm r}^{B}$. The teleported QFI is now given by
\begin{equation}
\mathcal{F}_{\phi}^{B}=\frac{4\sin^2\theta\overline{p}\ \overline{p}_{\rm r}^{B}\overline{\gamma}}{(N^{B})^2}.
\label{e24}
\end{equation}
The optimal post measurement strength $p_{\rm r}^{B}$ could be obtained by calculating the following conditions:
$\partial\mathcal{F}_{\phi}^{B}/\partial p_{\rm r}^{B}=0$, and $\partial^2\mathcal{F}_{\phi}^{B}/(\partial p_{\rm r}^{B})^{2}<0$.
The result turns to be
\begin{equation}
p_{\rm r}^{B,\rm opt}=1-\frac{\overline{p}\ \overline{\gamma}}{1+\overline{p}\gamma}, 
\label{e25}
\end{equation}
and the maximally teleported QFI is
\begin{equation}
\mathcal{F}_{\phi}^{B}=\frac{\sin^2\theta}{1+\overline{p}\gamma}.
\label{e26}
\end{equation}
Hence it can be seen clearly that the Eqs. (\ref{e25}) and (\ref{e26}) reduce to Eqs. (\ref{e16}) and (\ref{e17}) if we set $p=0$ (i.e., without the prior partial measurement). However, in contrast to scheme $A$, the most intriguing result is that the existence of prior partial measurement provides the possibility of dramatically enhancing the teleported QFI. As shown in Fig. \ref{Fig3}a, with the combination of prior and post partial measurement, the QFI could be greatly recovered by adjusting the partial measurement strength $p$. Particularly, when $p\rightarrow1$, $\mathcal{F}_{\phi}^{B}\rightarrow\sin^2\theta$ without regard to the strength of AD noise.

By comparing these two schemes, it is easy to conclude
that scheme $B$ is much more efficient than scheme $A$ on enhancing QFI teleportation. However, why does the scheme $B$ work much better than scheme $A$ with the assistance of a prior partial measurement? The underlying physics could be understood as follows: from Eq. (\ref{e5}), we know that the stronger the partial measurement strength $p$, the closer the initial EPR state is reversed towards the $|00\rangle$ state which is immune to AD noise. In the scheme $A$, no prior partial measurement is carried out before the qubit goes through the AD noise, thus the amount of reversed QFI highly depends on the decoherence strength $\gamma$. While in the second scheme, a prior partial measurement is performed to
move the state towards $|00\rangle$, which does not experience AD decoherence. Then an optimal partial measurement reversal is applied to revert the qubit back to the initial state. Therefore, the teleported QFI is not related to the decoherence strength $\gamma$ but depends on the prior partial measurement strength $p$. Full QFI can entirely be recovered by the combination prior partial measurement and post partial measurement reversal when $p\rightarrow1$.

The success probability of scheme $B$ is given by
\begin{equation}
P_{\rm QFI}^{B}=(1-p)(1-\gamma),
\label{e27}
\end{equation}
which obviously decays with the increase of $p$ and $\gamma$. Note that when $p\rightarrow1$, $P_{\rm QFI}^{B}\rightarrow0$. This means that the completely retrieve of QFI is attained at the expense of infinite low success probability. Nevertheless, the price of recovering QFI is still smaller than that of fidelity. As discussed in Ref. \cite{prama13}, the success probability of fidelity teleportation is 
\begin{equation}
P_{\rm fid}^{B}=\frac{\overline{\gamma}\ \overline{p}(2+\gamma\overline{p})}{2(1+\gamma\overline{p})}.
\label{e28}
\end{equation}
From Fig. \ref{Fig3}b, we note that $P_{\rm imp}^{B}\equiv P_{\rm QFI}^{B}-P_{\rm fid}^{B}\geqslant0$, which means the success probability of enhancing QFI is always higher than that of enhancing fidelity except for boundary values of $p$ and $\gamma$. This further confirms our conclusion that enhancing QFI is more economic than enhancing fidelity.

\subsection{Generalizations}
In the above analyses, we have assumed that Alice is not affected by noises.
In fact, these two schemes are universal for the case that both Alice and Bob suffer AD noise. Since scheme $A$ is a reduced version of scheme $B$, we only consider the later situation as an example. Here, Alice has to make prior partial measurements separately on Q1 and Q2 and then sends Q2 to Bob. It should be noted that Alice must act the post partial measurement before she does the Bell measurement while Bob should perform the post partial measurement after he has received the results sent by Alice. The finally teleported state can be characterized by the following Bloch vectors
\begin{eqnarray}
r_{x}&=&\frac{2\sin\theta\cos\phi}{N}\sqrt{\overline{p}_{1}\overline{p}_{2}\overline{\gamma}_{1}\overline{\gamma}_{2}\overline{p}_{\rm r_{1}}\overline{p}_{\rm r_{2}}},\\
\label{e29}
r_{y}&=&\frac{2\sin\theta\sin\phi}{N}\sqrt{\overline{p}_{1}\overline{p}_{2}\overline{\gamma}_{1}\overline{\gamma}_{2}\overline{p}_{\rm r_{1}}\overline{p}_{\rm r_{2}}},\\
\label{e30}
r_{z}&=&\frac{\cos\theta}{N}[\overline{p}_{\rm r_{1}}\overline{p}_{\rm r_{2}}+\overline{p}_{1}\overline{p}_{2}\overline{p}_{\rm r_{1}}\overline{p}_{\rm r_{2}}\gamma_{1}\gamma_{2}+\overline{p}_{1}\overline{p}_{2}\overline{\gamma}_{1}\overline{\gamma}_{2}\nonumber\\
&&-\overline{p}_{1}\overline{p}_{2}(\gamma_{1}\overline{\gamma}_{2}\overline{p}_{\rm r_{1}}+\overline{\gamma}_{1}\gamma_{2}\overline{p}_{\rm r_{2}})],
\label{e31}
\end{eqnarray}
with the normalized factor $N=\overline{p}_{\rm r_{1}}\overline{p}_{\rm r_{2}}+\overline{p}_{1}\overline{p}_{2}\overline{p}_{\rm r_{1}}\overline{p}_{\rm r_{2}}\gamma_{1}\gamma_{2}+\overline{p}_{1}\overline{p}_{2}\overline{\gamma}_{1}\overline{\gamma}_{2}+\overline{p}_{1}\overline{p}_{2}(\gamma_{1}\overline{\gamma}_{2}\overline{p}_{\rm r_{1}}+\overline{\gamma}_{1}\gamma_{2}\overline{p}_{\rm r_{2}})$. 
With these equations in hand, we can calculate the teleported QFI and optimize the variables $p_{\rm r_{1}}$ and  $p_{\rm r_{2}}$. In order to simplify the calculations, we assume that both qubit 1 and 2 interact with the same environments, i.e., $\gamma_1=\gamma_2=\gamma$. Consequently, we have $p_1=p_2=p$ and $p_{\rm r_{1}}=p_{\rm r_{2}}=p_{\rm r}$. The final results reduce to $p_{\rm r}^{\rm opt}=1-\overline{p}\ \overline{\gamma}/\sqrt{1+\overline{p}^2\gamma^2}$ and
\begin{equation}
\mathcal{F}_{\phi}^{\rm opt}=\frac{\sin^2\theta}{\left(\sqrt{1+\overline{p}^2\gamma^2}+\overline{p}\gamma\right)^2}.
\label{e32}
\end{equation}
From the above equation, we note that given the strength of AD noise $\gamma$, $\mathcal{F}_{\phi}^{\rm opt}$ achieves the minimum value with $p=0$. Namely, no prior partial measurement is performed, which  corresponds to the method of scheme $A$. Moreover, the minimum value of $\mathcal{F}_{\phi}^{\rm opt}$ is still larger than the pure AD noise case, which indicates that partial measurement indeed can be used for enhancing QFI teleportation even when both Alice and Bob are subject to AD noise. If $p\neq0$, i.e., prior partial measurements are carried out before the qubits undergo AD noise, the teleported QFI is further enhanced. Particularly, when $p\rightarrow1$, the initial QFI is almost entirely teleported to Bob.

\section{Discussions and Conclusions}
\label{sec:4}
Before conclusion, we point out that our proposals are entirely feasible with the present experimental techniques. The standard quantum teleportation has already been realized from photonic system to trapped ion and atomic systems. The techniques of partial measurement are also fast developed in recent years \cite{katz08,kim12,para06}. As demonstrated in Ref. \cite{kim12}, the partial measurements  can be implemented with a Brewster-angle glass plate (BAGP) for photon-polarization qubit because the BAGP probabilistically rejects vertical polarization ($|1\rangle$ state) and completely transmits horizontal polarization ($|0\rangle$ state), which exactly functions as the measurement depicted by Eqs. (\ref{e5}) and (\ref{e6}). On the other hand, the post weak measurement reversal could be realized by sequential bit-flip (a half-wave plate for polarization qubit), partial measurement and bit-flip. Alternatively, the partial measurements could be described with projective measurements in a larger Hilbert space that includes an ``ancilla qubit'' \cite{para11a,para11b}. Hence, the 
performance of partial measurement on the target qubit is equivalent to the action of von Neumann projective measurement on the ancilla qubit which is previously coupled to it.
Motivated by this guideline, partial measurements can be realized in any quantum system and not restricted to photon and superconducting qubits.

In summary, we propose to enhance the QFI teleportation under decoherence utilizing partial measurements. Thanks to the probabilistic nature of partial measurements, the teleported QFI could be greatly enhanced with the assistance of post partial measurement. Remarkably, we further show that the combined action of prior and post partial measurement can even completely keep QFI from AD noise. In addition, We demonstrate that the price of enhancing teleported QFI is smaller than that of improving fidelity.
Our work extends the ability of partial measurements as a new technique in various quantum information processing tasks, particularly, when the research objects are subject to AD noise.

\begin{acknowledgements}
This work is supported by the Funds of the National Natural Science
Foundation of China under Grant Nos. 11247006 and 11365011.
\end{acknowledgements}


\begin{thebibliography}{}

\bibitem{bennett93}C. H. Bennett, G. Brassard, C. Cr\'epeau, R. Jozsa, A. Peres, and W. K. Wootters, Phys. Rev. Lett. \textbf{70}, 1895 (1993).


\bibitem{gottesman99}D. Gottesman and I. Chuang, Nature (London) \textbf{402}, 390 (1999).

\bibitem{ursin04}R. Ursin \emph{et al}., Nature (London) \textbf{430}, 849 (2004).
\bibitem{braunstein98}S. L. Braunstein and  H. J. Kimble, Phys. Rev. Lett. \textbf{80}, 869 (1998).

\bibitem{bouwmeester97}D. Bouwmeester, J. W. Pan, K. Mattle, M. Eibl, H. Weinfurter, and A. Zeilinger, Nature (London) \textbf{390}, 575 (1997).
\bibitem{riebe04}R. Riebe \emph{et al}.,  Nature (London) \textbf{429} 734 (2004).
\bibitem{barret04}M. D. Barrett \emph{et al}., Nature (London) \textbf{429}, 737 (2004).

\bibitem{olms09}S. Olmschenk, D. N. Matsukevich, P. Maunz, D. Hayes, L. M. Duan, and C. Monroe, Science \textbf{323}, 486 (2009).

\bibitem{pfaff14}W. Pfaff \emph{et al}., Science \textbf{345}, 532 (2014).
\bibitem{wang15}X. L. Wang \emph{et al}., Nature \textbf{518} ,516 (2015).

\bibitem{takesue15}H. Takesue \emph{et al}., Optica \textbf{2}, 832 (2015).

\bibitem{helstrom76}C. W. Helstrom, \emph{Quantum Detection and Estimation Theory} (Academic, New York, 1976).

\bibitem{holevo82}A. S. Holevo, \emph{Probabilistic and Statistical Aspects of Quantum Theory} (North-Holland, Amsterdam, 1982).

\bibitem{kay93}S. M. Kay, \emph{Fundamentals of Statistical Signal Processing: Estimation Theory} (Prentice Hall, Upper Saddle River, 1993).


\bibitem{genoni11}M. G. Genoni, S. Olivares, and M. G. A. Paris, Phys. Rev. Lett. \textbf{106}, 153603 (2011).

\bibitem{lu13}X. M. Lu, Z. Sun, X. Wang, S. Luo, and C. H. Oh,
Phys. Rev. A \textbf{87}, 050302(R) (2013).

\bibitem{wootters81}W. K. Wootters, Phys. Rev. D \textbf{23}, 357 (1981).

\bibitem{nielsen00}M. Nielsen and I. Chuang, \emph{Quantum Computation and Quantum Information} (Cambridge University Press, Cambridge, 2000)

\bibitem{giov06}V. Giovannetti, S. Lloyd, and L. Maccone, Phys. Rev. Lett. \textbf{96}, 010401 (2006).

\bibitem{giov11}V. Giovannetti, S. Lloyd, and L. Maccone, Nat. Photon. \textbf{5}, 222 (2011).

\bibitem{braunstein94}S. L. Braunstein and C. M. Caves, Phys. Rev. Lett. \textbf{72}, 3439 (1994).


\bibitem{breuer1}H. P. Breuer and F. Petruccione, \emph{The Theory of Open Quantum Systems} (Oxford University Press, Oxford, 2002).
\bibitem{kolo10}J. Ko{\l}ody\'{n}ski and R. Demkowicz-Dobrza\'{n}ski, Phys. Rev. A \textbf{82} ,053804 (2010).

\bibitem{ma11}J. Ma, Y. X. Huang, X. Wang, and C. P. Sun, Phys. Rev. A \textbf{84}, 022302 (2011).

\bibitem{berrada12}K. Berrada, S. Abdel-Khalek, and A. S. Obada, Phys. Lett. A \textbf{376}, 1412 (2012).

\bibitem{zhang13}Y. M. Zhang, X. W. Li, W. Yang, and G. R. Jin, Phys. Rev. A \textbf{88}, 043832 (2013).

\bibitem{liyanling15}Y. L. Li, X. Xiao, and Y. Yao, Phys. Rev. A \textbf{91}, 052105 (2015).

\bibitem{korotkov99}A. N. Korotkov, Phys. Rev. B \textbf{60}, 5737 (1999).

\bibitem{korotkov06}A. N. Korotkov and A. N. Jordan, Phys. Rev. Lett. \textbf{97}, 166805 (2006).

\bibitem{korotkov10}A. N. Korotkov and K. Keane,  Phys. Rev. A \textbf{81}, 040103(R) (2010).


\bibitem{para11a}G. S. Paraoanu, Found. Phys. \textbf{41}, 1214 (2011).

\bibitem{para11b}G. S. Paraoanu, EPL  \textbf{93}, 64002 (2011).




\bibitem{sun09}Q. Q. Sun, M. Al-Amri, and M .S. Zubairy, Phys. Rev. A \textbf{80}, 033838 (2009).


\bibitem{man12}Z. X. Man, Y. J. Xia, and N. B. An,  Phys. Rev. A \textbf{86}, 012325 (2012).

\bibitem{xiao13}X. Xiao and Y. L. Li , Eur. Phys. J. D \textbf{67}, 204 (2013).


\bibitem{katz08}N. Katz \emph{et al}., Phys. Rev. Lett. \textbf{101}, 200401 (2008).

\bibitem{kim09}Y. S. Kim, Y. W. Cho, Y. S. Ra, and Y. H. Kim, Opt. Express \textbf{17}, 11978 (2009).
\bibitem{kim12}Y. S. Kim, J. C. Lee, O. Kwon, and Y. H. Kim, Nat. Phys. \textbf{8}, 117 (2012).


\bibitem{prama13}T. Pramanik and A. S. Majumdar, Phys. Lett. A \textbf{377}, 3209 (2013).

\bibitem{qiu14}L. Qiu, G. Tang, X. Yang, and A. Wang, Ann. Phys. (NewYork) \textbf{350}, 137 (2014).



\bibitem{yao14}Y. Yao, L. Ge, X. Xiao, X. Wang, and C. P. Sun, Phys. Rev. A \textbf{90}, 062113 (2014).
\bibitem{knysh11}S. Knysh, V. N. Smelyanskiy, and G. A. Durkin, Phys. Rev. A \textbf{83}, 021804(R) (2011).

\bibitem{liu13}J. Liu, X. Jing, and X. Wang, Phys. Rev. A \textbf{88}, 042316 (2013).  

\bibitem{zhong13}W. Zhong, Z. Sun, J. Ma, X. Wang, and F. Nori, Phys. Rev. A \textbf{87}, 022337 (2013).


\bibitem{wootters98}W. K. Wootters, Phsy. Rev. Lett. \textbf{80}, 2245 (1998).

\bibitem{para06}G. S. Paraoanu, Phys. Rev. Lett. \textbf{97}, 180406 (2006).








\end{thebibliography}


\end{document}